\documentclass{ws-ijmpd}

\begin{document}

\markboth{S D Maharaj and M Govender} {Radiating Collapse with
vanishing Weyl stresses}

%
\catchline{}{}{}{}{}
%

\title{Radiating Collapse with Vanishing Weyl stresses}

\author{S D Maharaj\footnote{E-mail: Maharaj@ukzn.ac.za}}

\author{M Govender \footnote{E-mail: Govenderm43@ukzn.ac.za}}

\address{School of Mathematical Sciences, \\
University of KwaZulu-Natal, Durban 4041, South Africa}

\maketitle

\begin{history}
\received{Day Month Year}
\revised{Day Month Year}
\end{history}

\begin{abstract}
In a recent approach in modelling a radiating relativistic star
undergoing gravitational collapse the role of the Weyl stresses
was emphasised. It is possible to generate a model which is
physically reasonable by approximately solving the junction
conditions at the boundary of the star. In this paper we
demonstrate that it is possible to solve the Einstein field
equations and the junction conditions exactly. This exact solution
contains the Friedmann dust solution as a limiting case. We
briefly consider the radiative transfer within the framework of
extended irreversible thermodynamics and show that relaxational
effects significantly alter the temperature profiles.
\end{abstract}


\section{Introduction}    

Since the pioneering work of Oppenheimer and Snyder\cite{snyder}
the final state of a star undergoing gravitational collapse has
occupied a significant standing in relativistic astrophysics. An
interesting extension of this problem is to include heat flow to a
spherically symmetric model as a realistic model of a radiating
star in gravitational collapse. Although this model is simplified,
it does provide a good insight into the nature of collapse and
serves as an excellent yardstick for more complicated radiating
models. The available literature\cite{bon}\cdash\cite{gov3} on
radiative gravitational collapse, and relativistic astrophysical
applications of spherically symmetric gravitational fields with an
imperfect matter distribution, is extensive. Models of
relativistic radiating stars are useful in the investigation of
the Cosmic Censorship hypothesis\cite{har}\cdash\cite{josh2} which
has attracted much attention in recent times amongst researchers
in the field.

Since the derivation of the junction conditions\cite{santos} for a
radiating spherically symmetric star undergoing gravitational
collapse and dissipating energy in the form of a radial heat flux,
there have been many novel approaches to finding physically viable
models. The model proposed by de Oliveira {\em et al.}
\cite{olive} in which the star collapses from an initial static
configuration has been studied extensively. Recently\cite{sailo}
it has been shown that a similar approach can be adopted to look
at the end state of collapse which results in a superdense star.
Another model which has been useful in understanding the effects
of dissipation is due to Kolassis\cite {kol} in which the fluid
trajectories are assumed to be geodesics. Recently Herrera {\em et
al} \cite {her1} have imposed the condition of conformal flatness
for a radiating, shear-free mass distribution undergoing
gravitational collapse. This approach generalises earlier works by
Patel and Tikekar \cite {lk} and Grammenos\cite{gram} in
particular. In their paper Herrera {\em et al.}\cite{her1}
provided an approximate solution of the Einstein field equations.
Their approximate solution highlighted the relaxational effects of
the collapsing fluid on the temperature profile. These results
were in agreement with earlier work carried out on the
thermodynamics of radiating stars within the
framework\cite{di1}\cdash\cite{gov3} of extended irreversible
thermodynamics. In this paper we demonstrate that it is possible
to solve the field equations exactly with the assumption of
vanishing Weyl stresses. Our exact solution provides a
mathematical basis for the approximate approach of Herrera {\em et
al.}\cite{her1} and indicates that their model is physically
significant.

This paper is organised as follows. In section two, we present the
field equations governing the interior spacetime comprising a
shear-free, spherically symmetric matter distribution with a
radial heat flux. We also impose the condition for conformal
flatness. In section three, the junction conditions required for
the smooth matching of the interior spacetime to the Vaidya
spacetime are outlined. In section four, an exact solution to the
boundary condition is presented. We show that our exact solution
has a Friedmann limit as in the case of the approximate analysis
of Herrera {\em et al.}\cite{her1} when the heat flux is absent.
The stability of the model is also briefly discussed. Section five
deals with the evolution of the temperature profiles within the
framework of extended irreversible thermodynamics. We are in a
position to calculate the temperature explicitly in both the
causal and noncausal theories and it is further shown that
relaxational effects lead to higher temperature within the stellar
core.

\section{Shear-free spacetimes}

We investigate the gravitational collapse of a shear-free matter
distribution with spherical symmetry. This is a reasonable
assumption when modelling a relativistic radiating star. In this
case there exists coordinates for which the line element may be
expressed in a form that is simultaneously isotropic and comoving.
With the coordinates $({x}^a ) = (t, r, \theta, \phi)$ the line
element for the interior spacetime of the stellar model takes the
form
\begin{equation}
ds^2 = -A^2dt^2 + B^2\left[dr^2 + r^2(d\theta^2 +
\sin^2{\theta}d\phi^2)\right] \,. \label{metric}
\end{equation}
where $A = A(t,r)$ and $B = B(t,r)$ are metric functions. In this
paper we consider a model which represents a spherically
symmetric, shear--free fluid configuration with heat conduction.
For our model the energy-momentum tensor for the stellar fluid
becomes
\begin{equation} T_{ab} = (\rho + p)u_a u_b + p g_{ab} + q_a u_b
          + q_b u_a \,.\label{2}
\end{equation} The fluid four--velocity ${\bf u}$ is comoving and is given by
\begin{equation} u^a = \displaystyle\frac{1}{A} \delta^{a}_0 \,.
\label{2'}
\end{equation} The heat flow vector takes the form
\begin{equation} q^a = (0, q, 0, 0)\,. \label{2''}
\end{equation} since $ q^au_a = 0 $ and the heat is assumed to flow in the
radial direction on physical grounds because of spherical
symmetry. The fluid collapse rate $\Theta = u^a_{;a}$ of the
stellar model is given by \begin{equation} \Theta =
3\frac{\dot{B}}{AB} \label{2'''}\end{equation} where dots
represent differentiation with respect to $t$. The nonzero
components of the Einstein's field equations for the line element
(\ref{metric}) and the energy-momentum (\ref{2}) are:
\begin{eqnarray}
\rho &=& \frac{3}{A^2}\frac{{\dot {B}}^2}{B^2} - \frac{1}{B^2}
\left( 2\frac{B''}{B} - \frac{{B'}^2}{B^2} +
\frac{4}{r}\frac{B'}{B} \right)  \label{14a} \\ \nonumber \\
p &=& \frac{1}{A^2} \left(-2\frac{{\ddot B}}{B} -\frac{{{\dot
B}}^2}{B^2} + 2\frac{{\dot A}}{A}\frac{{\dot {B}}}{B} \right)
\nonumber \\ \nonumber
 &&+ \frac{1}{B^2} \left(\frac{{B'}^2}{B^2} +
2\frac{A'}{A}\frac{B'}{B} + \frac{2}{r}\frac{A'}{A} +
\frac{2}{r}\frac{B'}{B} \right)  \label{14b}  \\  \nonumber \\
p &=& -2\frac{1}{A^2}\frac{{\ddot B}}{B} + 2\frac{{\dot
A}}{A^3}\frac{{\dot {B}}}{B} - \frac{1}{A^2}\frac{{\dot
B}^2}{B^2} + \frac{1}{r}\frac{A'{r}}{A}\frac{1}{B^2} \nonumber \\
\nonumber \\&& +  \frac{1}{r}\frac{B'}{B^3} +
\frac{A''}{A}\frac{1}{B^2} - \frac{{B'}^2}{B^4} + \frac{B''}{B^3}
\label{14c}  \\ \nonumber \\q &=& -\frac{2}{AB^2}
\left(-\frac{{\dot B}'}{B} + \frac{B'{\dot B}}{B^2} +
\frac{A'}{A}\frac{{\dot B}}{B} \right)\,. \label{14d}
\end{eqnarray} This is a system of coupled partial differential
equations in the variables $ A, B, \rho, p $ and $q$. The
nonvanishing components of the Weyl tensor $C_{abcd}$ for the line
element (\ref{metric}) are \begin{eqnarray} C_{2323} &=&
\frac{r^4}{3}B^2\sin^2{\theta}\left[\left(\frac{A'}{A} -
\frac{B'}{B}\right)\left(\frac{1}{r} + 2\frac{B'}{B}\right) -
\left(\frac{A''}{A} - \frac{B''}{B}\right)\right] \nonumber \\ \nonumber \\
C_{2323} &=&
-r^4\left(\frac{B}{A}\right)^2\sin^2{\theta}C_{0101}=2r^2\left(\frac{B}{A}\right)^2\sin^2{\theta}C_{0202}\nonumber
\\ \nonumber \\
&=& 2r^2\left(\frac{B}{A}\right)^2C_{0303}=
-2r^2\sin^2{\theta}C_{1212} = -2r^2C_{1313}\,. \end{eqnarray} If
we now impose the condition of conformal flatness, then this would
imply vanishing of all the Weyl tensor components. From the above
relations we note that this condition is fulfilled if we demand
\begin{equation} \label{weyl0} r\left(\frac{A''}{A} -
\frac{B''}{B}\right) - \left(\frac{A'}{A} -
\frac{B'}{B}\right)\left(1 + 2r\frac{B'}{B}\right) =
0\,.\end{equation} Equation (\ref{weyl0}) can be easily integrated
to give
\begin{equation}A(t,r) = [C_1(t)r^2 + 1]B\end{equation} where
$C_1(t)$ is an arbitrary function of $t$ yet to be determined.

The condition of pressure isotropy reduces to
\begin{equation}
\label{pi} \frac{B''}{B'} - 2\frac{B'}{B} - \frac{1}{r} =
0\end{equation} which has the general solution
\begin{equation} \label{pisol} B= \frac{1}{C_2(t)r^2 + C_3(t)}\end{equation}
where $C_2(t)$ and $C_3(t)$ are arbitrary functions of
integration.

\section{Junction Conditions}

Since the interior is radiating energy, the exterior spacetime is
described by Vaidya's outgoing solution given by\cite{vaidya}
\begin{equation} \label{v1}
ds^2 = - \left(1 - \frac{2m(v)}{R}\right) dv^2 - 2dvdR + R^2
\left(d\theta^2 + \sin^2\theta d\phi^2 \right)\,.
\end{equation}  The quantity $m$($v$) represents the Newtonian
mass of the gravitating body as measured by an observer at
infinity. The metric (\ref{v1}) is the unique spherically
symmetric solution of the Einstein field equations for radiation
in the form of a null fluid. The Einstein tensor for the line
element (\ref{v1}) is given by \begin{equation}  \label{v2} G_{ab}
= -\frac{2}{R^2}\frac{dm}{dv} \delta^0_a \delta^0_b
\,.\end{equation} The energy--momentum tensor for null radiation
assumes the form
\begin{equation} \label{v3} T_{ab} = {\Phi}w_{a}w_{b} \end{equation} where the null
four--vector is given by $w_a = (1, 0, 0, 0)$. Thus from
(\ref{v2}) and (\ref{v3}) we have that \begin{equation} \label{v4}
{\Phi} = -\frac{2}{R^2}\frac{dm}{dv}\end{equation} for the energy
density of the null radiation. Since the star is radiating energy
to the exterior spacetime we must have $\displaystyle\frac{dm}{dv}
\leq 0 $.

The necessary conditions for the smooth matching of the interior
spacetime to the exterior spacetime was first presented by
Santos\cite{santos} in his seminal paper. The junction conditions
for the line elements (\ref{metric}) and (\ref{v1}) are given by
\begin{eqnarray}
(rB)_{\Sigma} &=& R_{\Sigma}  \\
p_{\Sigma} &=& (qB)_{\Sigma}  \\
m(v) &=& \left\{\frac{r^3}{2}\left(\frac{{\dot B}^2B}{A^2} -
\frac{B'^2}{B}\right) - r^2B'\right\}_{\Sigma}\,.
\end{eqnarray}
Matching the interior spacetime (\ref{metric}) to the outgoing
Vaidya solution (\ref{v1}) leads to the following ordinary
differential equation
\begin{eqnarray} \label{111} &&\ddot{C_2}b^2 + \ddot{C_3} -
\frac{3}{2}\frac{(\dot{C_2}b^2 + \dot{C_3})^2}{(C_2b^2 + C_3)} -
\frac{\dot{C_1}b^2(\dot{C_2}b^2 + \dot{C_3})}{C_1b^2 + 1}
-2(\dot{C_3}C_1 - \dot{C_2})b \nonumber \\ \nonumber \\
&&+ 2\left(\frac{C_1b^2 + 1}{C_2b^2 + C_3}\right)\left[C_2(C_2 -
C_1C_3)b^2 + C_3(C_1C_3 - 2C_2)\right]= 0\,.\end{eqnarray} In the
above $C_1 = C_1(t)$, $C_2 = C_2(t)$ and $C_3 = C_3(t)$. Dots
denote differentiation with respect to $t$. Also $r_{\Sigma} = b$
is a constant which defines the boundary of the star. To complete
the gravitational description of this radiating star we must solve
the junction condition (\ref{111}).

\section{An exact model}

Herrera {\em et al.}\cite{her1} have presented a simple
approximate solution of (\ref{111}) which reduces to the Friedmann
dust sphere in the appropriate limit. They assumed the following
forms for the temporal functions \begin{equation} C_1 = \epsilon
C_1(t), \hspace{2cm} C_2 = 0, \hspace{2cm} C_3 =
\frac{a}{t^2}\,.\end{equation} where $0< \epsilon<< 1$ and $a$ is
a positive constant. With these assumptions, the junction
condition (\ref{111}) reduces to
\begin{equation}\label{app}
{\dot C_1} + \left(\frac{t}{b^2} + \frac{2}{b}\right)C_1 \approx
0\end{equation} which has the approximate solution
\begin{equation} \label{aapsol}
C_1 \approx C_1(0)\exp{\left(\frac{-t^2}{2b^2}
-\frac{2t}{b}\right)}\,.\end{equation} The approximate solution
(\ref{aapsol}) makes it possible to perform a qualitative analysis
of the physical features of the model.

However it is possible to solve (\ref{111}) exactly for particular
choices of the arbitrary temporal functions.  We now present a
simple exact solution of the junction condition (\ref{111}).
Guided by the approximate solution of Herrera {\em et
al.}\cite{her1} we take \begin{equation} \label{c2} C_2(t) = 0\,.
\end{equation} Substituting (\ref{c2}) into the boundary condition (\ref{111}) we obtain
\begin{equation} \label{linear}
C_3{\ddot C_3} - \frac{3}{2}{\dot C_3}^2 - \left[\frac{{\dot
C_1}b^2}{C_1b^2 + 1} + 2C_1b\right]C_3{\dot C_3} + 2(C_1b^2 +
1)C_1C_3^2 = 0\,.\end{equation} The transformation
\begin{equation} \label{u}
C_3(t) = u^{-2}
\end{equation}
enables us to write (\ref{linear}) in the equivalent form
\begin{equation} \label{master1}
{\ddot u} - {\dot u}\left[\frac{{\dot C_1}b^2}{C_1b^2 + 1} +
2C_1b\right]-(C_1b^2 + 1)C_1u = 0\,.\end{equation} Equation
(\ref{master1}) is a second order differential equation which is
linear in $u(t)$ if the function $C_1$ is specified. A simple
choice for $C_1$ is \begin{equation} \label{C1} C_1 = {\cal
C}\end{equation} where ${\cal C}$ is a constant. It follows that
(\ref{master1}) becomes
\begin{equation}
{\ddot u} - 2{\cal C}b{\dot u} - ({\cal C}b^2 +1){\cal C}u =
0\,.\end{equation}This differential equation has three categories
of solutions.

In terms of the original function $C_3$ we can present the
solutions as
\newline
{\bf Case I: \hspace{1cm} $2{\cal C}^2b^2 + {\cal C} >0$}
\newline
 \begin{equation} \label{I}C_3(t)
=\left[ \beta_1e^{({\cal C}b + \sqrt{2{\cal C}^2b^2 + {\cal
C}})t}+ \beta_2e^{\left({\cal C}b - \sqrt{2{\cal C}^2b^2 + {\cal
C}}\right)t}\right]^{-2}\,,\end{equation} {\bf Case
II:\hspace{1cm} $2{\cal C}^2b^2 + {\cal C} <0$}
\newline
\begin{equation} \label{II} C_3(t)= \left[\beta_1e^{2{\cal C}bt}\cos{(t\sqrt{2{\cal C}^2b^2 + {\cal C}})}
+\beta_2e^{{\cal C}bt}\sin{(t\sqrt{2{\cal C}^2b^2 + {\cal C}}
)}\right]^{-2}\,,\end{equation} {\bf Case III: \hspace{1cm}
$2{\cal C}^2b^2 + {\cal C} =0$}
\newline
\begin{equation} \label{III}
C_3(t)=(\beta_1 + \beta_2 t)^{-2}e^{{-2\cal C}bt}\,,\end{equation}
where $\beta_1$ and $\beta_2$ are constants of integration. We
have therefore exhibited an exact solution to the boundary
condition (\ref{111}) where
\begin{equation} \label{s}
C_1(t) = {\cal C}, \hspace{1cm} C_2(t) = 0\,,\end{equation} and
the three analytic forms for $C_3$ are given by
(\ref{I})-(\ref{III}). This exact solution provides a formal
mathematical basis for the approximate model of Herrera {\em et
al.}\cite{her1} for a radiating star.

We now consider {\bf Case III} more closely. The line element in
this case becomes
\begin{equation} \label{metric1}
ds^2 = {(\beta_1 + \beta_2t)^4}e^{4{\cal C}bt}\left[-\left({\cal C
}r^2 + 1\right)^2dt^2 + dr^2 + r^2(d\theta^2 +
\sin^2{\theta}d\phi^2)\right]\,. \label{1}
\end{equation}
If we set $\beta_1 = {\cal C} = 0, \beta_2 = 1$ in (\ref{metric1})
we obtain
\begin{equation} \label{metric2}
ds^2 = t^4\left[-dt^2 + dr^2 + r^2(d\theta^2 +
\sin^2{\theta}d\phi^2)\right] \label{1}
\end{equation}
which is the Friedmann dust solution. Herrera {\em et
al.}\cite{her1} also regain (\ref{metric2}) for their approximate
solution. However our limit arises from an exact model.

From (\ref{14a})-(\ref{14d}) we obtain the expressions
\begin{eqnarray}
\rho &=& \frac{12}{z^6}\frac{\left[\beta_2 + b{\cal
C}z\right]^2e^{-4{\cal C}bt}}{({\cal C}r^2 + 1)^2}
\\\nonumber\\
p&=& \frac{4{\cal C}}{z^5}\frac{\left[(1 - b^2{\cal C} + {\cal
C}r^2)\beta_1 + (-2b + t + (r^2 - b^2){\cal C}t )\beta_2\right]}{({\cal C}r^2 + 1)^2} \\ \nonumber \\
qB &=&-\frac{8{\cal C}r}{z^5}\frac{\left[\beta_2 + b{\cal
C}bz\right]e^{-4{\cal C}bt}}{({\cal C}r^2 + 1)^2}
\end{eqnarray}
for the line element (\ref{metric1}) where $z(t) = \beta_1 +
\beta_2 t$. The luminosity radius is given by
\begin{equation} \label{rad} (rB)_{\Sigma} = {(\beta_1 +
\beta_2t)^2}e^{2{\cal C}bt}\end{equation} which starts off at
arbitrary large values at $t = -\infty$ and evolves towards $t=0$.
The four-acceleration of this model is nonvanishing as
\begin{equation} \label{acc}
a_a = u^{b}\nabla_{b} u_a = \frac{A'}{A}\delta_a^1 = \frac{2{\cal
C}r}{1 +{\cal C}r^2}\end{equation} which is also independent of
time. The measure of the dynamical instability of the stellar
configuration at any given instant in time is given by
\begin{equation} \label{gamma}
\Gamma = \frac{d\ln{p}}{d\ln{\rho}}\,.\end{equation} For our model
we obtain
\begin{equation} \label{gam}
\Gamma_{centre} - \Gamma_{surface} = -\frac{b^2{\cal C}^2(\beta_1
+ \beta_2 t)^2}{6(\beta_2 + b{\cal C}(\beta_1 + \beta_2t)^2} <0\,.
\end{equation} This implies that
$\Gamma_{center}<\Gamma_{surface}$ for all times. This physical
result indicates that the centre of the collapsing star is more
unstable than the outer regions, reinforcing the approximate
results of Herrera {\em et al.}\cite{her1}  as well as the earlier
work by Denmat {\em et al.}\cite{den} which is also related.

\section{Thermodynamics}

In this section we investigate the evolution of the temperature
profile of our model within the context of extended irreversible
thermodynamics. The causal transport equation in the absence of
rotation and viscous stress is \begin{equation} \label{causalgen}
\tau h_a{}^b \dot{q}_b+q_a = -\kappa \left( h_a{}^b\nabla_b
T+T\dot{u}_a\right)
\end{equation} where $h_{ab}=g_{ab}+u_a
u_b$ projects into the comoving rest space, $T$ is the local
equilibrium temperature, $\kappa$ ($\geq0$) is the thermal
conductivity, and $\tau$ ($\geq 0$) is the relaxational time-scale
which gives rise to the causal and stable behaviour of the theory.
To obtain the noncausal Fourier heat transport equation we set
$\tau = 0$ in (\ref{causalgen}). For the metric (\ref{metric}),
equation (\ref{causalgen}) becomes \begin{equation} \label{causal}
\tau{(qB)}\!\raisebox{2mm}{$\cdot$}  + AqB = -\frac{\kappa
(AT)'}{B}\,.
\end{equation}
In order to obtain a physically reasonable stellar model we will
adopt the thermodynamic coefficients for radiative transfer. Hence
we are considering the situation where energy is transported away
from the stellar interior by massless particles, moving with long
mean free path through matter that is effectively in hydrodynamic
equilibrium, and that is dynamically dominant. Govender {\em et
al}\cite{gov1} have shown that the choice
\begin{equation} \kappa =\gamma T^3{\tau}_{\rm c}, \hspace{2cm}
 \tau_{\rm c} =\left({\alpha\over\gamma}\right)
T^{-\sigma}, \hspace{2cm}\tau =\left({\beta\gamma \over
\alpha}\right) \tau_{\rm c} \,, \label{5c}\end{equation} is a
physically reasonable choice for the thermal conductivity
$\kappa$, the mean collision time between massive and massless
particles $\tau_c$ and the relaxation time $\tau$. The quantities
$\alpha \geq 0$, $\beta \geq 0$ and $\sigma \geq 0$ are constants.
Note that the mean collision time decreases with growing
temperature as expected except for the special case $\sigma = 0$,
when it is constant.   With these assumptions the causal heat
transport equation (\ref{causal}) becomes
\begin{equation} \label{mm}
\beta(qB)\!\raisebox{2mm}{$\cdot$} T^{-\sigma} + A(qB) =
-\alpha\frac{T^{3 -\sigma}(AT)'}{B}\,.\end{equation} Solutions are
easily obtainable when in the noncausal case $(\beta = 0)$ as was
shown by Govinder and Govender\cite{gov3} in their general
treatment. All noncausal solutions of (\ref{mm}) are
\begin{eqnarray}
(A{\tilde T})^{4 - \sigma} &=& \frac{\sigma - 4}{\alpha}\int{A^{4
- \sigma}qB^2dr} + F(t), \hspace{2cm} \sigma \neq 0 \\ \nonumber \\
\ln{(A{\tilde T})} &=& -\frac{1}{\alpha}\int{qB^2 dr} + F(t),
\hspace{2cm} \sigma = 4\,. \end{eqnarray} where $F(t)$ is a
function of integration which is fixed by the surface temperature
of the star. Note that $\tilde {T}$ corresponds to the noncausal
temperature when $\beta = 0$. For a constant mean collision time
$(\sigma = 0)$, (\ref{mm}) can be integrated to give the causal
temperature, ie.,
\begin{equation}
(AT)^4 =
-\frac{4}{\alpha}\left[\beta\int{A^3B(qB)\!\raisebox{2mm}{$\cdot$}
dr} + \int{A^4qB^2dr}\right] + F(t)\,.\end{equation} In
(\ref{5c})we can think of $\beta$ as the `causality' index,
measuring the strength of relaxational effects, with $\beta=0$
giving the noncausal case.

The effective surface temperature of a star is given by
\begin{equation}
({\bar T}^4)_{\Sigma} =
\left(\frac{1}{r^2B^2}\right)\left(\frac{L}{4\pi\delta}\right)\end{equation}
where $L$ is the luminosity at infinity and $\delta (>0)$ is a
constant. The luminosity at infinity can be calculated from
\begin{equation}
L_{\infty} = -\frac{dm}{dv}\end{equation} where \begin{equation}
m(v) = \left[\frac{r^3B{\dot B}^2}{2A^2} - r^2B' -
\frac{r^3B'^2}{2B}\right]_{\Sigma}\,.\end{equation} For our model
we can calculate the temperature in both the causal and noncausal
theories explicitly. The noncausal temperature for constant
collision time is given by
\begin{equation} \label{nc}
{\tilde T}^4 = \left[\frac{1 + {\cal C}b^2}{1 + {\cal
C}r^2}\right]^4{\bar T}^4_{\Sigma} + \frac{16(\beta_2 z^{-1} +
{\cal C}b)}{3\alpha z^2}\left[\frac{(1 + {\cal C}r^2)^3 - (1 +
{\cal C}b^2)^3}{(1 + {\cal C}r^2)^4}\right]
\end{equation} where
$ z(t) = \beta_1 + \beta_2 t$. The causal temperature for constant
collision time is given by
\begin{eqnarray} \label{c}
{T}^4 &=& \left[\frac{1 + {\cal C}b^2}{1 + {\cal
C}r^2}\right]^4{\bar T}^4_{\Sigma} +
\frac{8\beta}{\alpha}\left[\frac{(\beta_2 z^{-1} + {\cal
C}b)}{z^4}e^{-4{\cal
C}bt}\right]\!\raisebox{3.5mm}{$\large\cdot$}\left[\frac{(1 +
{\cal C}r^2)^2 - (1 + {\cal C}b^2)^2}{(1 + {\cal C}r^2)^4}\right]
\nonumber \\&&+ \frac{16}{3\alpha}\frac{(\beta_2 z^{-1} + {\cal
C}b)}{z^2}\left[\frac{(1 + {\cal C}r^2)^3 - (1 + {\cal
C}b^2)^3}{(1 + {\cal C}r^2)^4}\right]e^{-2{\cal C}bt}\,.
\end{eqnarray} We note that at the boundary of the star, the
causal and noncausal temperatures, $T$ and $\tilde T$
respectively, are equal. In fact the causal temperature is
everywhere greater than its noncausal counterpart within the
interior of the star. From (\ref{causal}) we can write
\begin{equation} \label{ccc} \kappa(T)({A}T)^{\prime} -
\kappa({\tilde T})({A}{\tilde T})^{\prime} =
-(B\tau){(qB)}\!\raisebox{2mm}{$\cdot$}
\end{equation} which takes the form \begin{equation} \label{ccc1}
\left({\cal T}^{4 - \sigma}\right)^{\prime} - \left({\tilde{\cal
T}}^{4 - \sigma}\right)^{\prime} = - \frac{4 -
\sigma}{\alpha}\left({A}^{3 - \sigma} B
\tau\right){(qB)}\!\raisebox{2mm}{$\cdot$}
\end{equation} for the assumptions (\ref{5c}) where we have defined ${\cal T} = {A}T$ and
${\tilde{\cal T}} = {A}{\tilde T}$. We can immediately see that
the relative spatial gradient of ${\cal T}$ is greater than that
of ${\tilde {\cal T}}$, since $(qB)\!\raisebox{2mm}{$\cdot$} > 0$
and $4 - \sigma >0$. This result confirms earlier
findings\cite{gov1} in which the acceleration was vanishing and
the model had a Friedmann limit.

The evolution of the temperature has been studied within the
framework of extended irreversible thermodynamics and are in
keeping with earlier works on radiating collapse. This model has
shown that relaxational effects can significantly alter the
temperature of the star, especially during the latter stages of
collapse. We have further shown that the model is more unstable
closer to the centre as opposed to the outer regions of the star.
Such instablities have been shown to result in peeling or cracking
up\cite{her3} of the stellar configuration.

\end{document}